\documentclass[twocolumn,nofootinbib,superscriptaddress,amsmath,tightenlines,amssymb,aps,prd,preprintnumbers]{revtex4}

\newcommand{\R}{\ensuremath{\mathbb R}}
\newcommand{\C}{\ensuremath{\mathbb C}}
\newcommand{\Hf}{\ensuremath{\mathbb H}}
\newcommand{\Of}{\ensuremath{\mathbb O}}
\newcommand{\Z}{\ensuremath{\mathbb Z}}
\newcommand{\Q}{\ensuremath{\mathbb Q}}

\newcommand{\N}{{\mathcal N}}
\newcommand{\OO}{{\mathcal O}}
\newcommand{\HH}{{\mathcal H}}
\newcommand{\A}{{\mathcal A}}
\newcommand{\Top}{{\mathrm{top}}}

\newcommand{\abs}[1]{\lvert#1\rvert}

\newcommand{\pa}{{\partial}}

\newcommand{\fg}{{\mathfrak g}}
\newcommand{\fp}{{\mathfrak p}}

\newcommand{\fh}{{\mathfrak h}}

\newcommand{\ti}[1]{\textit{#1}}
\newcommand{\tzeta}{\tilde\zeta}

\newcommand{\ii}{{\mathrm i}}

\begin{document}

\title{BPS black holes, quantum attractor flows and automorphic forms}

\author{Murat Gunaydin}
\affiliation{Department of Physics, Penn State University, University Park, PA 16802, USA}
\email{murat@phys.psu.edu}

\author{Andrew Neitzke}
\affiliation{School of Natural Sciences, Institute for Advanced Study, Princeton, NJ 08540, USA}
\email{neitzke@ias.edu}

\author{Boris Pioline}
\affiliation{LPTHE, Universit{\'e}s Paris VI \& VII, Jussieu, 75252 Paris, FRANCE}
\affiliation{LPTENS, D{\'e}partement de Physique de l'ENS,
24, rue Lhomond, 75005  Paris, FRANCE}
\email{pioline@lpthe.jussieu.fr}

\author{Andrew Waldron}
\affiliation{Department of Mathematics, University of
             California, Davis, CA 95616, USA}
\email{wally@math.ucdavis.edu}

\date{December 22, 2005; Revised on March 24.}

\preprint{LPTENS-05-42, {\tt hep-th/0512296}}

\begin{abstract}
We propose a program for counting microstates of four-dimensional
BPS black holes in~$\N \geq 2$ supergravities
with symmetric-space valued
scalars by exploiting the symmetries of timelike reduction to three
dimensions. Inspired by the equivalence between the
four dimensional attractor flow
and geodesic flow on the three-dimensional scalar manifold,
we radially quantize stationary, spherically symmetric
BPS geometries. Connections
between the topological string amplitude, attractor wave function,
the Ooguri-Strominger-Vafa conjecture and the theory of automorphic forms
suggest that black hole degeneracies are counted by Fourier coefficients
of modular forms for the three-dimensional U-duality group, associated
to special ``unipotent'' representations which appear in the
supersymmetric Hilbert space of the quantum attractor flow.
\end{abstract}

\maketitle

\section{Introduction}

Understanding the microscopic origin of the thermodynamic entropy of
black holes is a challenge for any quantum theory of gravity.  In
string theory a microscopic description of certain supersymmetric
black holes is known and accounts for the expected number of microstates in
the limit of large charges~$Q \gg 1$
(\cite{Strominger:1996sh, Maldacena:1996gb} and much subsequent work).
Recently there has been some progress in understanding
the counting of microstates beyond the strict large $Q$ limit
(see {\it e.g.}~\cite{deWit:2005ya} for a recent review).

Strikingly, Ooguri, Strominger and Vafa (OSV)~\cite{Ooguri:2004zv}
have argued that, in
$\N=2$ compactifications of Type II string theory on a Calabi-Yau
threefold~$X$, all subleading~$1/Q$
corrections to the degeneracies should be computable
by considering the squared
topological string amplitude~$\abs{\Psi_\Top}^2$ as a generating
function at fixed magnetic charge.
A possible rationale for this phenomenon
uses channel duality in the
near horizon~$AdS_2\times S^2$~\cite{Ooguri:2005vr}: the partition function with
compactified Euclidean time, which counts black hole microstates
(possibly with signs), could equivalently be computed in radial quantization
as an overlap of wave functions at each
boundary of~$AdS_2$. Provided the
topological string amplitude solves
the radial Wheeler-DeWitt constraint, the general form of the
OSV conjecture then follows in analogy with open/closed string duality
on the cylinder. This equivalence, if correct, may be viewed as a
mini-superspace AdS/CFT correspondence, where the bulk
evolution of the gravity fields is related to time evolution in a
(as yet unknown) dual conformal quantum mechanics.

On the other hand, in theories with high~$\N=4$ or
$\N=8$ supersymmetry, the exact black hole degeneracies
have long been suspected
to be related to Fourier coefficients of some kind of automorphic
form~\cite{Dijkgraaf:1996it}.  Indeed, the respective
four-dimensional U-duality groups~$G_4=SL(2,\Z) \times SO(6,n,\Z)$ or
$E_{7(7)}(\Z)$ provide strong constraints on microscopic
degeneracies (see~\cite{Obers:1998fb} for a U-duality review).
Recently however, there have been indications that the {\it
three}-dimensional U-duality groups~$G_3= SO(8,n+2,\Z)$ or
$E_{8(8)}(\Z)$ may be spectrum generating symmetries for {\it
four}-dimensional BPS black holes
\cite{Ferrara:1997uz,Gunaydin:2000xr,Gunaydin:2005gd,Pioline:2005vi}.
This idea is partly motivated by the ``quasi-conformal'' realization
of~$G_3$~\cite{Gunaydin:2000xr} on the space of electric and
magnetic charges~$(p^I,q_I)$ extended by one
variable.  This is parallel to  the ``conformal'' realization of~$G_4$ on the
space of black hole charges in five dimensions
\cite{Ferrara:1997uz,Gunaydin:2000xr,Gunaydin:2005gd}. Upon
quantization, it leads to the minimal unipotent representation of
$G_3$~\cite{Gunaydin:2000xr,Kazhdan:2001nx}, which is a crucial
ingredient in the construction of a particular automorphic form for
$G_3$.

In this Letter, we outline a new approach to
black hole microstate counting inspired by these ideas,
postponing details to~\cite{gnpw-to-appear}.
Taking the proposal in~\cite{Ooguri:2005vr} literally,
we study the radial quantization of stationary,
spherically symmetric BPS black holes
in four-dimensional~$\N\geq 2$ supergravity theories. To avoid
the complications of full-fledged~$\N=2$ supergravity, we restrict to
special cases in which the scalars take values in a symmetric space
$G_4/K_4$.  For simplicity, we further discard any higher-derivative
corrections to the low energy effective action.
This allows us to jettison Calabi-Yau geometry
in favor of real Lie group representation theory, while retaining
some essential features of a realistic~$\N=2$ supergravity. Higher
$\N=4$ or~$\N=8$ supergravities are also in this category, provided we
again discard higher-derivative corrections.

Since our interest lies in stationary solutions, it is natural to
perform a reduction along the time
direction, leading to three-dimensional Euclidean gravity coupled to a
non-linear sigma
model with an enlarged symmetry group~$G_3$~\cite{Breitenlohner:1987dg}.
An essential
observation is that stationary, spherically symmetric solutions of
the four-dimensional equations of motion are equivalent to geodesic
trajectories on the three-dimensional scalar manifold
$M_3^*=G_3/H_3$, where~$H_3$ is a certain noncompact real form of
the maximal compact subgroup~$K_3$. As such, they carry an action of
$G_3$ relating black hole geometries with different charges\footnote{We note
that $G_3$ already made an appearance as a solution-generating symmetry
in the $\N=4$ context \cite{Cvetic:1995kv}.}.

This observation is key to the radial quantization of black holes in a
mini-superspace approximation.  Geodesic trajectories can be replaced
by square-integrable wave functions on $M_3^*$ (or sections of some
bundle). The action of $G_3$ is then roughly the regular
representation on the Hilbert space $L^2(G_3/H_3)$, which
breaks up into a sum of unitary irreducible representations (irreps) of
$G_3$. As we shall see, the BPS component of the Hilbert space is a
``unipotent'' representation of unusually small functional
dimension. Furthermore, in some cases there is a distinguished ``spherical''
vector in this space, which appears to play the role of a
1-parameter extension of the topological string amplitude
$\Psi_\Top$, in broad accordance with the proposal
of~\cite{Ooguri:2005vr}.

\begin{figure*}
\begin{tabular}{|c|c|c|c|c|} \hline
$n_Q$ & $n_v$ & $M_4$           & $M_3^*$  & $J$  \\ \hline \hline
    8        & 1  & $\emptyset $  & $U(2,1) / [U(1,1)\times U(1)]$ & $\R$                 \\ \hline
    8        & 2  & $SL(2,\R) / U(1)$  & $G_{2,2} / SO(2,2)$ & $\R$                 \\ \hline
    8        & 7  & $Sp(6,\R) / (SU(3) \times U(1))$ & $F_{4(4)} / [Sp(6,\R) \times SL(2,\R)]$  &  $J_3^\R$ \\ \hline
    8        & 10 & $SU(3,3) / (SU(3) \times SU(3) \times U(1))$ & $E_{6(2)} /
[SU(3,3) \times SL(2,\R)]$ &  $J_3^\C$ \\ \hline
    8        & 16 & $SO^*(12) / (SU(6) \times U(1))$ & $E_{7(-5)} /
[SO^*(12) \times SL(2,\R)]$ &  $J_3^\Hf$ \\ \hline
    8        & 28 & $E_{7(-25)} / (E_{6} \times U(1))$ & $E_{8(-24)} / [E_{7(-25)} \times SL(2,\R)]$ &  $J_3^\Of $\\ \hline
    8        &  $n+2$  & $(SL(2,\R) / U(1)) \times (SO(n,2)/[SO(n) \times SO(2)])$ & $SO(n+2,4) / [SO(n,2) \times SO(2,2)]$ &  $\R \oplus \Gamma_{n-1,1} $
\\ \hline \hline
   16        &  $n+2$  & $(SL(2,\R) / U(1)) \times (SO(n-4,6)/[SO(n-4) \times SO(6)])$ & $SO(n-2,8) / [SO(n-4,2) \times SO(2,6)]$ &  $\R \oplus \Gamma_{n-5,5} $
\\ \hline \hline
   32        & 28 & $E_{7(7)} / SU(8) $ & $E_{8(8)} / SO^*(16)$ &  $J_3^{\Of_s}$ \\ \hline
\end{tabular}
\caption{Number of supercharges, vector fields in~$d=4$,
moduli spaces before and after~$d=4 \to 3$ reduction
along a timelike Killing vector, and associated Jordan algebra.  Real
forms of exceptional groups are indicated by the notation~$E_{n(m)}$
where~$m$ is the difference between the number of non-compact
and compact generators.}
\end{figure*}

In the full quantum theory, pursuing the suggestion in~\cite{Pioline:2005vi},
we conjecture that a discrete subgroup~$G_3(\Z)$
remains as a ``spectrum generating'' symmetry controlling the
exact degeneracies of BPS black holes.  This idea is supported by
the fact that the above unipotent representations have been
exploited by mathematicians to construct modular forms for~$G_3$.
Thus it is tempting to identify the black hole degeneracies as particular
generalized Fourier coefficients of these modular forms. As
we shall see, the result~\eqref{four} below bears a close resemblance to the OSV
conjecture, albeit with striking differences.

The proposal that BPS microstates are counted by
modular forms for~$G_3$ should have interesting ramifications (if it
can be verified).  For example, it could provide an explanation of
mysterious modular properties which have been observed in computations
of the degeneracies of BPS states in four
dimensions~\cite{Dijkgraaf:1996it, Aganagic:2004js,Vafa:2004qa,
Gaiotto:2005gf,Shih:2005qf,Pioline:2005vi,Shih:2005uc}.
More generally, we are hopeful that understanding the
spectrum generating symmetries of black hole
degeneracies will improve our understanding of the conjecture
of~\cite{Ooguri:2004zv} and the non-perturbative meaning of the
topological string.

Finally, it should be mentioned that the radial quantization of
black holes has been a subject of much work by the canonical gravity
community~\cite{Thiemann:1992jj,Kuchar:1994zk,Cavaglia:1994yc,Hollmann:1996cb,Hollmann:1996ra,Breitenlohner:1998yt}. While this endeavor
has so far yielded relatively little insight into black hole
entropy, embedding this idea in supergravity
may justify the mini-superspace truncation, at least for BPS black
holes, and allow channel duality to teach us the nature of
black hole microstates.

The organization of this Letter is as follows: in Section II, we describe
a class of supergravities with symmetric scalar manifolds where our
techniques apply most directly. In Section III, we interpret the
radial evolution equations as geodesic motion on a scalar manifold
$M_3^*$ and determine the conditions on geodesics
required for supersymmetric solutions. In Section IV, we explain our proposal
for quantizing the attractor flow, and argue that the BPS Hilbert space
furnishes a particular unipotent representation of~$G_3$. In Section V,
we conjecture a relation between exact black hole degeneracies and
Fourier coefficients of automorphic forms naturally associated to
these representations.  We close in Section VI with an outlook.

\section{Homogeneous supergravities}

The analysis we describe applies to a class of~$d=4$, $\N=2$
supergravity theories whose scalar fields are associated to vector
multiplets and lie on a symmetric space~$M_4 = G_4 / K_4$.  Such
theories, known as ``very special supergravities'', were first
studied in ~\cite{Gunaydin:1983rk,Gunaydin:1983bi,Gunaydin:1984ak}.
The special geometry of $M_4$
turns out to be characterized by a cubic prepotential~$F = N(X)
/ X^0$, with~$N(X)$ the norm function of a degree 3 Jordan algebra
$J$. The classification and study of such theories is therefore
closely tied to the theory of Jordan algebras.

We also consider theories with higher supersymmetry,~$\N=4$ or
even~$\N=8$.  Here again one finds scalar fields living on some
symmetric space~$M_4 = G_4 / K_4$.  For example, in the~$\N=8$ case (the low
energy limit of M-theory compactified on~$T^7$) this space is~$E_{7(7)}
/ (SU(8) / \Z_2)$.
We list some examples in the Table, giving the number of supercharges,
the moduli space in~$d=4$, and the moduli space in~$d=3$ after
reduction along a timelike Killing vector (see the next Section). For
brevity we omit discrete factors such as~$\Z_2$. We shall refer to
these~$\N\geq 2$ supergravities with symmetric-space-valued scalars
as ``homogeneous supergravities''.

Passing to the quantum theory, the continuous symmetry is generally
broken to a discrete subgroup (the continuous symmetry
would be inconsistent with charge quantization).  In the~$\N=8$ case
it is believed that the quantum theory is invariant under the
largest subgroup consistent with this constraint, written
$E_{7(7)}(\Z)$.  For the very special supergravity theories the
quantum theory is not in general known to exist; we
optimistically assume that the quantum theory exists and
is invariant under a suitably large discrete subgroup
$G_4(\Z)$\footnote{ Of the four~$\N=2$ supergravity theories
defined by simple Jordan algebras of degree three, those with 7, 10
and 16 vector fields in four dimensions can be obtained by a
consistent truncation of the maximal~$\N=8$ supergravity. The theory
with 10 vector fields is known to describe the vector-multiplet sector
of type IIA  superstring compactified on the CY orbifold
$T^6/\Z_3$~\cite{Ferrara:1988fr}.  There are indications that
the 6 extra scalars in the model based on $J_3^\Hf$ can be considered
as generalized complex structure moduli \cite{Pestun:2005ni}}. It is conceivable, but
currently unproven, that our considerations extend to
general non-homogeneous~$\N=2$ theories, the monodromy group of the
Calabi-Yau periods playing the role of~$G_4(\Z)$.

\section{Attractor equations and geodesic flow}

Since we are interested in stationary black
hole solutions it is natural to dimensionally reduce the~$d=4$ theory
along a timelike direction, using a Kaluza-Klein-type
ansatz~\cite{Breitenlohner:1987dg}
\begin{equation}
\label{ans4}
ds_4^2 = - e^{2U} (dt + \omega)^2 + e^{-2U} ds_3^2\ .
\end{equation}
This leads to Euclidean gravity in three
dimensions, coupled to scalars, vectors and fermions; upon dualizing
the bosonic sector is described simply by the three-dimensional
metric~$ds_3^2$ and scalar fields~$\phi^a$.
The~$\phi^a$ include the scalars from the
$d=4$ theory, plus electric and magnetic potentials from the reduction of
the gauge fields~$A^I_t$ and their duals, plus the scale factor~$U$ and
the twist potential dual to the shift~$\omega$;
altogether they are organized into a manifold
$M_3^*$ of indefinite signature
(the analytic continuation of the
Riemannian signature manifold~$M_3$ from a spacelike reduction).
For generic~$\N=2$ theories,~$M_3$ is a~$4n_v$ dimensional
quaternionic-K\"ahler manifold
known as the c-map of~$M_4$~\cite{Ferrara:1989ik};
$M_3^*$ is an associated para-quaternionic space. For homogeneous
supergravities,~$M_3=G_3 / K_3$ and~$M_3^*=G_3 / H_3$, where
$K_3$ is the maximal compact subgroup of~$G_3$ and~$H_3$ is a non-compact
real form of~$K_3$~\cite{Breitenlohner:1987dg,Hull:1998br}.

Stationary black hole configurations are identified with solutions of
the non-linear sigma model on~$M_3^*$ coupled to three-dimensional gravity.
Supersymmetry requires a flat three-dimensional slicing, so that
general BPS solutions are harmonic maps from
$\R^3$ to~$M_3^*$. Hence the problem of constructing black hole
geometries, and possibly also counting their microstates,
is related to the study of such harmonic maps.

As a first step, let us restrict to spherically symmetric configurations,
whose metric on three-dimensional slices may be written
\begin{equation}
ds^2_3 = N^2(\rho)\, d\rho^2 + r^2(\rho)\left[ d\theta^2 + \sin^2 \theta\, d \phi^2 \right]\,.
\end{equation}
Considering~$\rho$ as a ``radial time'' the bosonic action reads
\begin{equation}
\label{sbos}
S = \int d\rho \left[ \frac{N}{2} + \frac{1}{2N}
\left( \dot{r}^2 - r^2 G_{ab} \dot{\phi}^a \dot{\phi}^b \right) \right],
\end{equation}
where~$\phi: \R \to M_3^*$ describes the radial scalar field evolution
and~$G_{ab}$ is the metric on~$M_3^*$.  The 4-dimensional
equations of motion are thus equivalent to geodesic motion of a fiducial
particle on the real cone over~$M_3^*$. The equation of motion for the
lapse~$N$ imposes the Hamiltonian constraint
\begin{equation}
H = p_r^2 - \frac{1}{r^2} G^{ab} p_a p_b - 1 \equiv 0\, ,
\end{equation}
(where~$p_r$ and~$p_a$ are the canonical conjugates to~$r$ and
$\phi^a$) which fixes the mass of the fiducial particle on the cone
to~1. For BPS black holes, one may set~$N=1, \rho=r, p_r=1$ so that
the problem reduces to {\it light-like} geodesic motion on~$M_3^*$,
with affine parameter~$\tau=1/r$. It is convenient to retain the~$r$
variable to define observables such as the horizon area and the ADM
mass.

The magnetic and electric charges of the black hole
can be easily read off from this description:
they are Noether charges~$p^I$,~$q_I$ associated to the generators of
gauge transformations in the isometry group~$G_3$ acting on
$M_3^*$.  These charges do not commute; rather, they generate a
Heisenberg subgroup~$N \subset G_3$:
\begin{equation}
\label{heis}
[p^I, q_J]_{\rm PB} = 2 \delta^I_J\  k\, ,
\end{equation}
where~$k$ is the NUT charge of the black hole~\cite{Pioline:2005vi}.
Configurations with~$k\neq 0$ have closed timelike curves at infinity
when lifted back to four
dimensions, due to the off-diagonal term~$\omega=k\cos\theta d\phi$
in the metric~\eqref{ans4};
{\it bona fide} four-dimensional black holes are only obtained
in the ``classical'' limit~$k\to 0$. Nonetheless, it is advantageous
to retain this variable to realize
the full three-dimensional U-duality symmetry.
In addition, the isometry
associated to rescalings of~$g_{tt}$ leads to an additional conserved
charge~$m$ identified as the ADM mass, obeying
\begin{equation}
[m,p^I]_{\rm PB}=p^I,\ [m,q_I]_{\rm PB}=q_I,\ [m,k]_{\rm PB}=2k\, .
\end{equation}
Importantly, as already noted in~\cite{Hollmann:1996cb}, the ADM mass does
not Poisson-commute with the charges.

For homogeneous supergravities where~$M_3^*=G_3/H_3$, there exist
additional conserved charges associated to the isometric action of
$G_3$ on~$M_3^*$. The full symmetry group~$G_3$ includes the group
$G_4$ already present in 4 dimensions, the unipotent subgroup~$N$,
and its opposite~$\bar N$, corresponding to generalized Ehlers and Harrison
transformations~\cite{kinnersley,Breitenlohner:1987dg}.
The corresponding set of conserved charges can be
assembled into an element~$Q$ in the Lie algebra dual~$\fg_3^*$ of~$G_3$
(which we canonically identify with $\fg_3$ via the Killing form) ---
the moment map of the symplectic action of
$G_3$ on phase space. The space of charges~$Q$ naturally breaks up
into orbits of various dimensions under the (co)adjoint action of~$G_3$,
all equipped with a canonical Kirillov-Kostant
symplectic structure.
As we shall now see, the supersymmetry properties of the black hole
solution are simply expressed in terms of the conjugacy class of
$Q$.

While the flatness of the three-dimensional slices is  necessary
for supersymmetry, it is not a sufficient condition. For spherically
symmetric stationary solutions,
the supersymmetry variation of the fermionic partners
$\lambda^A$ of~$\phi^a$ reads~\cite{Bagger:1983tt}
\begin{equation}
\label{dla}
\delta\lambda^A = V^A_\alpha \epsilon^\alpha,
\end{equation}
where~$\epsilon^\alpha$ is the supersymmetry parameter and~$V^A_\alpha$
is a matrix linear in the velocities~$\dot\phi^a$ on~$M_3^*$.
For general~$\N=2$ supergravities, the
indices~$\alpha=1,2$ and~$A=1,\dots, 2n_v$ transform as fundamental
representations
of the restricted
holonomy group~$Sp(2,\R)\times Sp(2n_v,\R)$ of para-quaternionic geometry.
Supersymmetry is preserved when~\eqref{dla} vanishes
for some non-zero~$\epsilon^\alpha$, which implies
that the quaternionic viel-bein~$V^A_\alpha$ has a zero eigenvector.
Expressing~$V^A_\alpha$ in
terms of the conserved charges, one may show that this
amounts to the system of equations
\begin{eqnarray}
\label{genatt1}
\frac{dz^i}{d\tau} &=& - e^{U+i\alpha} g^{i\bar j} \partial_{\bar j} |Z| \\
\label{genatt2}
\frac{dU}{d\tau} + \frac{i}{2} k &=& - 2 e^{U+i\alpha} |Z|\, ,
\end{eqnarray}
where
\begin{equation}
Z(p,q,k) = e^{K/2} \left[ (q_I - 2 k \tzeta_I) X^I - (p^I + 2k\zeta^I) F_I \right]
\end{equation}
is the central charge of the supersymmetry algebra,
and the phase~$\alpha$ is adjusted so that
$dU/dt$ is real. The scalars $\tzeta_I,\zeta^I$, conjugate to the
charges~$p^I$,~$q_I$, themselves evolve according to
\begin{eqnarray}
\frac{d\zeta^I}{d\tau} &=&
-\frac12 e^{2U} [(\Im {\cal N})^{-1}]^{IJ} \nonumber\\
&&\left[ q_I - 2 k \tzeta_I - [\Re {\cal N}]_{JK} ( p^K + 2 k \zeta^K)
\right]\\
\frac{d\tzeta_I}{d\tau} &=&
-\frac12 e^{2U} [\Im {\cal N}]_{IJ} ( p^J + 2 k \zeta^I)
-[\Re {\cal N}]_{IJ}
\frac{d\zeta^J}{d\tau} \nonumber
\end{eqnarray}
where ${\cal N}_{IJ}$ is the standard matrix of special geometry~(we
follow the conventions from~\cite{Ceresole:1995ca}).
For vanishing NUT charge\footnote{The attractor equations including NUT charge
can also be obtained as a special case of the analysis in \cite{Denef:2000nb};
they agree with ours, although the proof is not
immediate~\cite{gnpw-to-appear}.} ~$k=0$,
Eqs~\eqref{genatt1},\eqref{genatt2}
are recognized as the standard attractor flow equations describing the
radial evolution of the scalars toward the black hole
horizon~\cite{Ferrara:1995ih,Ferrara:1996um,Ferrara:1997tw,
Moore:1998pn}.
The isomorphism between attractor flow on the special K\"ahler
manifold~$M_4$ and supersymmetric geodesic motion on the c-map of
$M_4$ was in fact independently noticed in~\cite{Gutperle:2000ve} in their
study of spherically symmetric D-instantons in five-dimensional
$\N=2$ supergravity, and reflects mirror symmetry between
D-instantons and black holes.

For homogeneous~$\N=2$ supergravity, the holonomy group is further restricted
to~$G_4 \times Sp(2,\R)$, where~$G_4 \subset Sp(2n_v,\R)$.
The matrix of velocities~$V_\alpha^A$ can be viewed as an element of
$\fp$ in the decomposition~$\fg_3= \fh \oplus \fp$, and is
conjugate to the matrix of conserved charges~$Q$. The condition for
supersymmetry is then\footnote{One can see this by noting that~$\fg_3$ has a 5-grading
induced by the decomposition under a highest root of~$Sp(2,\R)$, and arguing
that~$V_1^A\propto V_2^A$ can be conjugated into the grade-1 subspace.}
\begin{equation}
[Ad(Q)]^5=0\ ,
\end{equation}
where~$Ad(Q)$ denotes the adjoint representation of~$Q$. In other words,
$Q$ should be an element of a \ti{nilpotent orbit}~$\OO_5$.
The dimension of this orbit is~$4n_v+2$, much smaller than the
dimension~$8n_v$ of the unconstrained phase space.
One may also consider yet
smaller orbits~$\OO_4, \OO_3, \OO_2$, of
dimensions\footnote{The fact
that the dimension of~$\OO_3$ is an even integer can be traced back to the Jordan
algebra origin of homogeneous supergravities.}~$4n_v, (10n_v-4)/3,
2n_v+2$, corresponding to 3-charge, 2-charge and 1-charge black holes,
with zero entropy in tree-level supergravity.
All of these preserve the same amount of supersymmetry, but belong
to different duality orbits. The minimal orbit~$\OO_2$ will play
an important role in the relation to the topological amplitude below.

This discussion may be repeated for theories with higher supersymmetry:
for~$\N=8$, the fermionic variation is~$\delta\lambda_A = \epsilon_\alpha
\Gamma^\alpha_{A\dot A} P^{\dot A}$ where~$\epsilon_\alpha, \lambda^A,
P^{\dot A}$ transform as a vector, spinor and conjugate spinor  of the
R-symmetry group~$SO^*(16)$, respectively. A
1/2-BPS black hole in~$\N=8$ preserving 16 supercharges
is obtained when the velocity~$P^{\dot A}$ is a pure spinor of~$SO^*(16)$,
equivalently~$Q\in\OO_2$: this corresponds to the minimal orbit of~$E_8$,
of dimension 58. Similarly, a 1/4-BPS black hole
corresponds to~$Q\in\OO_3$, of dimension
$92$, while a generic 1/8-BPS black hole
corresponds to~$Q\in\OO_5$, of dimension~$114$. The orbit~$\OO_4$
of dimension 112 corresponds to 1/8-BPS black holes with
zero classical entropy, but does not enjoy any enhanced
supersymmetry~\cite{Ferrara:1997uz}.
Note that the dimensions of these orbits agree with those given
above for~$n_v=28$.

\section{The quantum attractor flow}

Having reduced the radial evolution of stationary, spherically symmetric
solutions in 4 dimensions to geodesic flow on the three-dimensional
scalar manifold~$M_3^*$, quantization is straightforward:
Classical geodesic motion of a particle on~$\R^+ \times M_3^*$
is replaced by
a wave function~$\Psi \in L^2(\R^+ \times M_3^*)$, satisfying the quantum
Hamiltonian constraint
\begin{equation}
\left[-\frac{\pa^2}{\pa r^2} + \frac{\Delta}{r^2} -1 \right]\,
\Psi(r,e) = 0\, ,
\end{equation}
where~$\Delta$ is the Laplace-Beltrami operator on~$M_3^* \ni e$.
In our case we are really dealing with the motion of a superparticle, so~$\Psi$ should
more properly be a section of a bundle over~$\R^+ \times M_3^*$.

Furthermore, we are interested in the BPS Hilbert space, composed of
states satisfying a quantum version of~\eqref{dla} and~$p_r=1$.  The
latter equation takes care of the radial dependence,~$\Psi(r,e) = r
\Psi(e)$. We shall discard the~$r$ variable from now on.

Let us first sketch the quantization of~\eqref{dla} for general~$\N=2$
supergravities. Note first that (after continuing from~$M_3^*$ to
$M_3$)~$\epsilon^\alpha$ determines a complex
structure at a point of $M_3$.  Classically, the relation~\eqref{dla}
implies that the
holomorphic part of the velocity~$V^A_\alpha$ for this complex
structure vanishes.  We propose that quantum mechanically this should
translate into the statement that~$\Psi$ belongs to a holomorphic
sheaf cohomology group~$H^1(T,{\mathcal O}(-h))$ on the
twistor space\footnote{It may be fruitful to further extend $T$
to the Hyperk\"ahler cone over $M_3$, along the lines of~\cite{deWit:2001dj,
Rocek:2005ij}. We thank M.~Rocek and S.~Vandoren for discussions on this
approach.}
$T$ of the quaternionic-K\"ahler space $M_3$.
(In particular,~$\Psi$ depends holomorphically on~$2n_v+1$
variables, giving the expected functional dimension for the BPS
Hilbert space.)  So we are proposing a relation between sections of
bundles on~$M_3^*$ and sheaf cohomology on~$T$; this is likely related
to the higher-dimensional quaternionic version of the Penrose
transform, discussed in~\cite{quatman,MR1165872}.

For homogeneous~$\N \geq 2$ supergravities, this construction can be
made more concrete as follows.  We impose constraints on the phase
space by fixing the values of all of the Casimir operators of~$G_3$
acting on the Noether charge.  After imposing these constraints, the
phase space reduces to a union of coadjoint orbits of the Noether
charges.  In particular, if we impose the constraint (following from
the BPS condition) that the Casimir operators all vanish, the reduced
phase space is the union of nilpotent orbits.  Correspondingly, the
BPS Hilbert space~${\cal H}_{BPS}$ decomposes into a direct sum of
irreducible unitary representations of~$G_3$,\footnote{Since~$G_3$ is
a non-compact group, its non-trivial unitary representations are
infinite-dimensional; nevertheless they may be characterized by their
\ti{functional} (``Gelfand-Kirillov'') dimension, the number
of variables needed to represent them.} obtained by quantizing these
nilpotent coadjoint
orbits\footnote{To be precise, the above construction provides a
quantization of the observables associated to the action of~$G_3$.
Fortunately, the most relevant observables~$p^I,q_I,k,m$ are indeed
generators of~$G_3$. Other observables need not act within~${\cal
H}_{BPS}$, although they may be projected into it in the spirit of the
lowest Landau level truncation in condensed matter physics.}.  The
orbits relevant for us are the~$\OO_i$ discussed in the previous
Section, corresponding to duality orbits of the black hole charges.

Each representation $(\rho, \HH)$ obtained by orbit quantization can be thought of
as embedded in the unconstrained Hilbert space of sections on
$G_3/H_3$: namely, there is a ``spherical'' $H_3$-covariant\footnote{More precisely, the vector
$f$ transforms in a finite-dimensional representation of~$H_3$ appropriate to the
bundle over~$G_3/H_3$; it is a spherical vector if this representation is trivial.}
vector~$f \in \HH$, so given some $v \in \HH$, we define a section $\Psi_v(g)$ by
\begin{equation}
\label{mapsph}
\Psi_{v}(g) = \langle f, \rho(g) v \rangle\, .
\end{equation}
Note that~$\Psi_{v}(g)$ is well defined as a section on~$G_3/H_3$ thanks to the
$H_3$-covariance of~$f$.

So the radial quantization of supersymmetric stationary spherically
symmetric geometries, or supersymmetric geodesic motion on~$M_3^*$, is
equivalent to the quantization of the small nilpotent orbits~$\OO_i$
of~$G_3$, where~$i=2,3,4,5$ depending on the fraction of supersymmetry
preserved by the solution (or, in the~$\N=2$ case, on the number of
independent charges carried by the black hole).  This is a problem
which has been considered at some length in the mathematics and physics
literature.  In particular, the minimal representation, based on the
quantization of~$\OO_2$, was constructed for all
simply laced Lie groups in their split real
form in~\cite{MR1159103,MR1278630}, while
the small representations~$\HH_{2,3,4,5}$ associated to~$\OO_{2,3,4,5}$ were
constructed in~\cite{MR1421947} for all simple Lie groups in their
quaternionic real form. Their construction of~$\OO_5$
is in fact a particular case of the twistor construction for general~$\N=2$
supergravity sketched above.
The quantization of~$\OO_5$ in the split case was also physically
realized as a quasi-conformal action leaving a quartic light-cone
invariant in~\cite{Gunaydin:2000xr}; the
quantization of~$\OO_2$ was obtained in~\cite{Gunaydin:2001bt} by
``quantizing the quasi-conformal action''; these constructions were
extended to the non-split case in~\cite{Gunaydin:2004md,Gunaydin:2005zz}. The
same construction
of the minimal representation in the split case was also independently
arrived at in~\cite{Kazhdan:2001nx}, albeit with very different physical
motivations. In that work the distinguished spherical vector was also
computed.

At this stage, we may try to make contact
with the proposal of~\cite{Ooguri:2005vr}: are any of the~$\HH_i$
suitable Hilbert spaces to contain the topological string amplitude
$\Psi_\Top$?  Taking~\cite{Ooguri:2005vr} literally, the
desired wave function must be a function of~$n_v$
variables.  Alas, there is no unitary representation
of~$G_3$ on this number of variables, but the minimal representation
comes very close: it has functional dimension~$n_v + 1$.
Furthermore, choosing an electric-magnetic splitting yields
an explicit realization of the minimal representation as a function
space~\cite{Kazhdan:2001nx};~$n_v$ of the variables can be naturally
identified with the magnetic charges~$p^I$ of the four-dimensional black
hole, while the last one can be identified with the NUT
charge~$k$~\cite{Pioline:2005vi}.
Modulo the extra variable, this is exactly the form expected for the
topological string partition function, including the dependence on a
choice of splitting.

Moreover, there is a natural way to eliminate
the extra variable: letting~$H_\omega$ denote the generator conjugate
to the ADM mass, one may consider the limit~$\lim_{\tau \to \infty}
e^{\tau H_\omega} \Psi$ (which may be thought of as a ``4-dimensional''
or perhaps ``near-horizon'' limit).  In this limit,~$G_3$ is broken down to
$G_4 \times \R$ commuting with~$H_\omega$, and the spherical vector
computed in~\cite{Kazhdan:2001nx} reduces to the tree-level generalized
topological string amplitude~$\Psi_\Top = e^{N(X)/X^0}$. It is thus
tempting to interpret the topological string amplitude as a restriction
of the spherical vector\footnote{In particular, it is not accurate to say
that~$\Psi_\Top$ is the wave function for the attractor flow: rather, (a
one-parameter generalization of) it defines a map~\eqref{mapsph}
from~${\cal H}_{BPS}$ to the unconstrained Hilbert space; the wave function itself is
determined only when a vector~$v \in {\cal H}_{BPS}$ is specified. In
the next section, we propose a principle that selects a unique~$v$.}
of the minimal representation for the three-dimensional
duality group~$G_3$. It would obviously be very
interesting to have a topological string interpretation of the extra
variable~$k$, and conversely to understand how perturbative and instanton
corrections modify the notion of spherical vector.

\section{The automorphic wave function}

In the last Section we discussed four unipotent representations~$\HH_i$
of the three-dimensional continuous U-duality group~$G_3$, which arise upon
quantization of spherically symmetric BPS black hole attractor flows.
Mathematical interest in these representations
lies in the fact that they allow for the construction of simple modular
forms for~$G_3$; so we begin by reviewing this notion and then explain its
relevance for us.

Recall that an automorphic form for a group~$G$ is
a function on the quotient~$G(\Z)\backslash G$, where~$G(\Z)$ is
a discrete subgroup of~$G$ (see {\it e.g.}~\cite{MR0207650} for a more precise
definition).  The space~$\A$ of automorphic forms has a natural $G$-action
(by right multiplication).
A modular form~$\Psi$ is abstractly defined as an equivariant map
$\HH \to \A$ where the space~$\HH$ carries a representation~$\rho$ of~$G$:
$\HH$ characterizes the ``modular weight'' of~$\Psi$, while the map
encodes the precise modular form.  In other words,~$\Psi$ denotes a particular
realization of the representation~$\rho$ on a space of functions on~$G(\Z) \backslash G$.
One way to construct such a
realization\footnote{See {\it e.g.}~\cite{Pioline:2003bk} for a physicist's
introduction
to this approach.} is
to find a~$G(\Z)$-invariant distribution~$f_{G(\Z)}$ in~$\HH^*$: then
the map
\begin{equation}
\label{modu}
v \mapsto \Psi_v(g) =  \langle f_{G(\Z)}, \rho(g) v \rangle
\end{equation}
defines a modular form. If the representation~$\rho$ admits a
vector~$v_K$ invariant under~$K\subset G$, the resulting function
$\Psi_{v_K}$ is~$K$-invariant, hence a function on the double
coset~$G(\Z)\backslash G/K$.  In some cases the
invariant distribution~$f_{G(\Z)}$ can itself be obtained as an adelic
spherical vector, expressible as the product over all primes~$p$ of
spherical vectors over $G(\Q_p)$, where~$\Q_p$ is the
field of~$p$-adic rational numbers.

We briefly indicate how the usual modular forms for~$SL(2,\R)$ fit into
this framework.  Holomorphic modular forms~$f(\tau)$ on the upper half-plane give
functions on $SL(2,\Z) \backslash SL(2,\R)$ by\footnote{As usual, for $g = \begin{pmatrix}a & b \\ c & d \end{pmatrix}$, $(f \vert_k g)(\tau) = (c \tau + d)^{k} f\left(\frac{a \tau +b}{c \tau + d}\right).$} $g \mapsto (f \vert_k g)(\ii)$.  Under the action of $SL(2,\R)$ this is a highest weight vector generating the~$k$-th holomorphic discrete series representation.
The Jacobi theta series also fits in this scheme via~\eqref{modu},
upon choosing~$\HH$ as the metaplectic
representation,
$v_K(x)$ as the ground state of the harmonic oscillator, and~$f_{G(\Z)}$
the ``Dirac comb'' distribution~$\sum_{m\in\Z} \delta(x-m)$.

Now recall that given a modular form for~$SL(2,\R)$ its Fourier coefficients
often have interesting number-theoretic properties ---
they are integers and answer counting problems.
For example, for holomorphic modular forms of~$SL(2,\Z)$, one obtains these coefficients by computing
\begin{equation}
\label{foursl2}
\hat\Psi(m)
= \int_0^1 f(\tau_1 + \ii \tau_2) e^{-2\pi\ii m (\tau_1 + \ii \tau_2)} d\tau_1\, .
\end{equation}
This can be equivalently viewed as integration over the parabolic
subgroup~$\left\{ \begin{pmatrix} 1 & x \\ 0 & 1
\end{pmatrix} \right\} \subset SL(2,\R)$ fixing the cusp~$\tau = \ii
\infty$, modulo the action of~$SL(2,\Z)$.
In favorable cases, a similar prescription can be given for Fourier
coefficients of modular forms of~$G$,
by performing integrals over the unipotent (``upper
triangular'') parts of suitable parabolic subgroups; this procedure
is not known to work in full generality, but does
at least for the~$\N=2$ cases of interest
here~\cite{MR1932327,MR1988198,d4-mod}.

Namely, let us consider modular
forms~$\Psi$ for~$G = G_3$, the
three-dimensional U-duality group appearing in the timelike reduction
of stationary black holes, taking~$\rho$ to be one of the unipotent representations
$\HH_2, \HH_3, \HH_5$ which are singled out by supersymmetry.
This~$G_3$ has a maximal parabolic subgroup~$G_3 = LN$, where the L\'evi component
$L \simeq G_4 \times \R$ contains the four-dimensional U-duality along
with~$H_{\omega}$, and~$N$ is the Heisenberg group
of gauge transformations from~\eqref{heis}.  Given~$\Psi$
one can try to extract its Fourier
coefficients.  The analog of~$e^{2 \pi \ii n x}$ in
\eqref{foursl2} is now a character~$\chi: N \to \C^\times$, with~$\chi(n) =
1$ for~$n \in G_3(\Z)$.  Such characters are naturally parameterized
by vectors~$(p^I, q_I)$ of integrally quantized electric and magnetic
black hole charges, the NUT charge~$k$ being necessarily set to 0 (there
exists a different unipotent subgroup of~$G_3$ whose characters are
parameterized by~$q_I$ and~$k$, but then the~$p^I$'s necessarily vanish).

The Fourier coefficients~$\hat \Psi(p,q)$ so obtained
are by construction invariant under four-dimensional U-duality
$G_4(\Z)$; importantly they are constrained to fit together into a
modular form for a larger group~$G_3(\Z)$. Furthermore, the
``smallness'' of the underlying representation implies
that only a subset of the Fourier coefficients are nonzero,
namely those lying in the appropriate orbit of the U-duality group.

In particular, let us consider the minimal representation~$\HH_2$.  In
this case it is believed that the modular form~$\Psi$ is unique.
So our conjecture in this case would be that the Fourier coefficients
of this~$\Psi$ count the microstates of 1-charge black holes.  Although these
microstate counts are not of much direct physical interest, the
statement that they fit together into a modular form of~$G_3$ is an
illustration of our general philosophy.

When~$G_3$ is a split group, the modular form attached to the
minimal representation can be understood concretely by
considering the spherical vector; its corresponding
automorphic form was constructed in~\cite{minspher}, essentially using the
formula~\eqref{modu}.  Comparing~\eqref{modu}
to~\eqref{mapsph}, we see that this construction corresponds to
a particular choice of wave function~$v=f_{G(\Z)}$ in the corresponding
BPS Hilbert space, which we refer to as the ``automorphic wave function''.
Then the Fourier coefficient associated to the given charges~$(p^I, q_I)$ can be formally
expressed as
\begin{equation}
\label{four}
\hat \Psi(p,q) =  \int d\zeta^I\  e^{i q_I \zeta^I}\
f^*_{G(\Z)}( p^I - \zeta^I, 0) \,
f_{K(\R)}( p^I + \zeta^I, 0) \ .
\end{equation}
This expression strongly resembles the OSV conjecture but differs by
involving the product of the real and adelic spherical vectors,
rather than the squared modulus of the topological string amplitude.

For the other~$\HH_i$, coming from quantization of the larger~$\OO_i$,
the situation becomes more complicated; the
modular forms here are not expected to be unique, so some additional
input will be required to pin down the desired ones.  The OSV conjecture
seems to suggest considering the tensor product of the
representation~$\HH_2$ with itself and projecting onto~$\HH_5$; but
this is unlikely to be the full story, since from the results
of~\cite{Dijkgraaf:2005bp} it appears that considering non-spherically
symmetric
configurations requires the inclusion of higher powers of~$\Psi_\Top$
as well.

\vspace*{4mm}

\section{Summary and outlook}

In this work, motivated by recent conjectures about exact
degeneracies of BPS black holes in four dimensions, we studied
stationary, spherically symmetric solutions of~$\N\geq 2$
supergravities, with emphasis on cases where the scalar manifold is
a symmetric space. By utilizing the equivalence with geodesic motion
on the three-dimensional scalar manifold, we quantized the radial
attractor flow, and argued that the three-dimensional U-duality
group~$G_3$ acts as a spectrum generating symmetry for BPS black
hole degeneracies in 4 dimensions. We suggested how these may be
counted by Fourier coefficients of modular forms of~$G_3$. Clearly
much work remains to be done in this direction. Other outstanding
problems are to understand the r\^ole of rotating and multi-centered
black holes, as well as the effect of higher derivative corrections,
and to understand the appearance of the extra NUT-charge parameter~$k$
in the generalized topological string amplitude. It should also be
pointed out that a similar line of reasoning may be developed for
5-dimensional black holes and black rings, leading us to expect that
the 4-dimensional U-duality group plays a similar role in~5
dimensions. Finally, the extension of this approach to full-fledged
$\N=2$ string theory, if successful, is likely to uncover new
relations between number theory, Calabi-Yau geometry and physics.

\begin{acknowledgments}

We thank Iosif Bena, Frederik Denef, Ori Ganor, Eric Gimon, Benedict Gross,
Simeon Hellerman, Soji Kaneyuki, Erez Lapid, Jian-Shu Li, Juan
Maldacena, Stephen Miller, Hirosi Ooguri, Martin Ro\v{c}ek,
Peter Sarnak, Lior Silberman, Stefan
Vandoren, David Vogan, Martin Weissman, and Edward Witten for very
valuable discussions and explanations. M.~G. and B.~P. are grateful
to KITP for hospitality during part of this work, and to the organizers
of the program ``Mathematical Structures in String Theory'' for a stimulating
workshop.  The research of
A.~N. is supported by the Martin A. and Helen Chooljian Membership at the
Institute for Advanced Study.  This research was
supported in part by the National Science Foundation under Grant Nos.
PHY99-0794, PHY-0245337, PHY-01-40365 and PHY-0503584.

\end{acknowledgments}

\renewcommand{\href}[2]{{#2}}

\bibliography{aut-short}

\end{document}